\documentclass{article}
\usepackage[utf8]{inputenc}
\usepackage{geometry}
\usepackage{graphicx}
\usepackage{xcolor}
\usepackage{bm}
\usepackage{amsmath}
\usepackage{adjustbox}
\usepackage[normalem]{ulem}
\usepackage{graphicx}
\usepackage{url}
\usepackage{color,soul}
\usepackage{gensymb}
\usepackage[numbers,square]{natbib}
\usepackage{notoccite}
\usepackage{authblk}
\usepackage{mathtools}

\begin{document}
\title{\textbf{Active Learning A Neural Network Model For Gold Clusters \& Bulk From Sparse First Principles Training Data}}
\author[1]{Dr. Troy D Loeffler$^*$}
\author[1,2]{Dr. Sukriti Manna\thanks{These two authors contributed equally}}
\author[3]{Dr. Tarak K Patra}
\author[1,2]{\\ Dr. Henry Chan}
\author[4]{Dr. Badri Narayanan}
\author[1,2]{Dr. Subramanian Sankaranarayanan\thanks{Corresponding author: skrssank@uic.edu}}
\affil[1]{Center for Nanoscale Materials, Argonne National Laboratory \\ Lemont, Illinois 60439, United States}
\affil[2]{Department of Mechanical and Industrial Engineering, University of Illinois \\ Chicago, Illinois 60607, United States}
\affil[3]{Department of Chemical Engineering, Indian Institute of Technology Madras \\ Chennai, TN 600036, India}
\affil[4]{Department of Mechanical Engineering, University of Louisville, Louisville \\ KY 40292, USA}

% \date{\today}

\maketitle
\begin{abstract}
Small metal clusters are of fundamental scientific interest and of tremendous significance in catalysis. These nanoscale clusters display diverse geometries and structural motifs depending on the cluster size; a knowledge of this size-dependent structural motifs and their dynamical evolution has been of longstanding interest. Given the high computational cost of first-principles calculations, molecular modeling and atomistic simulations such as molecular dynamics (MD) has proven to be an important complementary tool to aid this understanding. Classical MD typically employ predefined functional forms which limits their ability to capture such complex size-dependent structural and dynamical transformation. Neural Network (NN) based potentials represent flexible alternatives and in principle, well-trained NN potentials can provide high level of flexibility, transferability and accuracy on-par with the reference model used for training. A major challenge, however, is that NN models are interpolative and  requires large quantities ($\sim 10^4$ or greater) of training data to ensure that the model adequately samples the energy landscape both near and far-from-equilibrium. A highly desirable goal is minimize the number of training data, especially if the underlying reference model is first-principles based and hence expensive. Here, we introduce an active learning (AL) scheme that trains a NN model on-the-fly with minimal amount of first-principles based training data. Our AL workflow is initiated with a sparse training dataset ($\sim $ 1 to 5 data points) and is updated on-the-fly via a Nested Ensemble Monte Carlo scheme that iteratively queries the energy landscape in regions of failure and updates the training pool to improve the network performance. Using a representative system of gold clusters, we demonstrate that our AL workflow can train a NN with $\sim$ 500 total reference calculations. Using an extensive DFT test set of $\sim{~}$1100 configurations, we show that our AL-NN is able to accurately predict both the DFT energies and the forces for clusters of a myriad of different sizes. Our NN predictions are within 30 meV/atom and 40 meV/\AA \space of the reference DFT calculations. Moreover, our AL-NN model also adequately captures the various size-dependent structural and dynamical properties of gold clusters in excellent agreement with DFT calculations and available experiments. We finally show that our AL-NN model also captures bulk properties reasonably well, even though they were not included in the training data.
\end{abstract}

\maketitle
\section{\label{sec:one}Introduction}
Small clusters approaching the sub-nanometer size range have attracted a lot of interest in catalytic applications.~\cite{liu2018metal, hamilton1979catalysis} Such size-selected clusters that comprise of a handful of atoms often display exotic catalytic properties that are much different than that of either nano-sized or bulk catalysts.~\cite{tyo2015catalysis,kawawaki2020gold,jans2019gold, yamazoe2014nonscalable,li2013atomically,sak2019sustainable,
sengupta2010recent} 
These clusters contain well-defined number of atoms and offer an ideal platform to study catalysis at the atomic level. They also serve as model systems to enable a comprehensive fundamental insight into the nature of the catalytic processes that are otherwise difficult to explore using catalysts prepared by conventional methods that often yield particles with finite distributions in size and composition.

The recent advances in synthesis science has allowed us to exercise precise control over the structure and composition of these small catalytic clusters. For example, Vajda and coworkers have shown that sub-nanometer Pt clusters can serve as highly active and highly selective catalyst for the oxidative dehydrogenation of propane.~\cite{vajda2009subnanometre} More recently, they have also shown that sub-nanometer sized cobalt oxide clusters can enable oxidative dehydrogenation of cyclohexane at lower temperatures than conventional catalysts, while eliminating the combustion channel.~\cite{tyo2012oxidative} These individual clusters that contain a handful of atoms have a high surface-to-volume ratio and much higher fraction of undercoordinated atoms. Apart from displaying exceptional catalytic activity, they also offer an excellent and economic utilization of the metal loading.~\cite{guo2014pt} In view of these studies, an area of growing interest is to design new catalytic materials in an atom-by-atom fashion.
A lot of catalyst design work focuses on exploring conditions and pathways for their synthesis and are effectively aimed at tuning the number of under-coordinated sites via experimental controls such as pressure, temperature etc. From this perspective, physically accurate, flexible and accurate MD simulations and models are important to enable insilico design given the exhaustive space that needs to be explored and the experimental trials being time-consuming and costly. The recent advances in computational resources and first-principles based methods have allowed for rapid high-throughput computational studies to design catalysts.~\cite{greeley2006computational,tosoni2019oxide, strasser2003high} More recently, the advances in data science and machine learning have allowed for computations to provide a better characterization to complement experiments and extract more information about the structure and compositions of these catalyst.~\cite{kitchin2018machine} Computations based on density functional theory have allowed us to uniquely explore the energetics and thermodynamics of high-energy intermediates or transient metastable states that play an important role in the catalytic pathway that may escape experimental characterization.~\cite{alavi1998co, anstrom2003density, norskov2011density} 

Apart from energetics, the dynamical evolution of these clusters is also important from a design perspective. These clusters undergo dynamical processes that involve structural transitions from one stable/metastable state to another; often these metastable states have been shown to display much higher catalytic activity than their stable counterparts.~\cite{hostetler1999dynamics} Ab-initio molecular dynamics (AIMD) techniques represents a popular method to probe the dynamics. But despite the improvements in computational resources, the AIMD simulations are limited in the timescales and length-scales that they can access. Furthermore, it is also worth noting that the the global minimum energy configurations of these catalytic clusters in the mid-size regime ($n=$ 20-100) are not well understood. Such exhaustive structural searches for these sizes remain intractable within the framework of high-fidelity calculations such as DFT even with the most efficient sampling methods (e.g., evolutionary algorithms,~\cite{huang2013ground, zhao2016comprehensive} basin-hopping,~\cite{zhan2005asynchronous, ouyang2015global, kim2008new} etc.).

A classical description of the potential energy surface of these small clusters can provide a cheaper surrogate to perform either longer time dynamical simulations or carry out an exhaustive search of the structure/compositional space of these catalytic clusters. The primary challenge with these models is that they trade accuracy for computational efficiency. Despite being popular, classical models with pre-defined functional forms struggle to accurately describe the structure and dynamics of clusters in the ($n$= 10-100) range. For instance, Au clusters in the sub-nanometer range undergo a planar-to-globular transition at cluster size of 13 atoms, which has proven to be very difficult for empirical potential models to capture.~\cite{kinaci2016unraveling} Spherically symmetric potentials such as embedded atom method and Sutton Chen potentials cannot capture the planar configurations whereas bond-order potentials such as Tersoff perform well for planar structures but do not completely capture the size dependent structural transition in Au clusters.~\cite{narayanan2016describing} It is well known that the use of predefined functional form imposes serious limitations on the physics and chemistry that can be captured. 

Neural network (NN) based potential models offer a flexible alternative to capture the size dependent structural and dynamical transformations in these nano and sub-nanoscale catalysts.~\cite{chiriki2017neural, jindal2017spherical, ouyang2015global, artrith2014understanding} Recently, NN models are emerging as a popular technique due to the rapid advancement in the computational resources as well as the myriad of electronic structure codes that allow for efficient generation of the training data.~\cite{behler2011neural} The underlying goal in the development of these NN models is to train against vast amounts of high-fidelity first-principles data and thereby replicate their accuracy at a fraction of their computational cost. An inherent limitation of these NN models is that they are interpolative and as such the traditional approach for training a NN has often relied on generating as large a training data as is possible. Such large-scale generation of high-fidelity training data can become challenging depending on the level of the electronic structure calculations employed.~\cite{pilania2017multi}

To address the issue with training data generation, there have been several recent efforts to device active learning strategies that allow for efficient sampling of training data for NN models. Smith et al. employed an active learning (AL) strategy based on the Query by Committee (QBC) scheme.\cite{smith2018less} QBC uses the disagreement between an ensemble of ML potentials to infer the reliability of the ensemble's prediction. QBC allowed for automatic sampling of the regions of chemical space where the potential energy was not accurately described by the ML potential. Their AL approach was validated on a test set consisting of a diverse set of organic molecules and their results showed that one requires only 10\% to 25\% of the data to accurately represent the chemical space of these molecules. 

Similarly, Zhang et al.~\cite{zhang2019active} introduced an AL scheme (deep potential generator (DP-GEN)) that constructs ML models for simulating materials at the molecular scale. Their procedure involve exploration, generation of accurate reference data, and training. They used Al and Al-Mg as representative cases and showed that ML models can be trained with minimum number of reference data.  In another work, Vandermause et al.~\cite{vandermause2019accelerating} sampled structures on-the-fly from AIMD and used an adaptive Bayesian inference method to automate the training of low-dimensional multiple element interatomic force fields. Their AL framework uses internal uncertainty of a Gaussian process regression model to decide acceptance of model prediction or the need to augment training data. In all of the above studies, the overarching aim in these studies is to minimize the ab-initio training data required to describe the potential energy surface.

Here, we introduce a new active learning (AL) strategy~\cite{loeffler2020active,loeffler2020active2,patra2019coarse} that learns the potential energy surface description from minimal amount of first-principles training data sampled from on-the-fly Monte Carlo simulations. Our workflow starts with minimal training data ($\sim$ 1 to 5 data points) and is continually updated via a Nested Ensemble Monte Carlo scheme that iteratively queries the energy landscape of the catalyst in regions of failure and improves the network performance. 

We choose a model catalysts system i.e. gold clusters and deploy the AL strategy to develop a NN potential model that can accurately predict the diverse geometries of gold clusters. For the small clusters, when bond-directionality effects are important, popular spherically symmetric potential models such as embedded atom method (EAM)~\cite{foiles1986embedded,daw1984embedded} or its variants like Sutton-Chen (SC)~\cite{doye1998global} and Gupta potentials,~\cite{shao2005structural} cannot capture size dependent transitions and the diverse cluster geometries. Bond-order based empirical force fields (EFF) (e.g, Tersoff-type BOP,~\cite{backman2012bond} Reactive force field ReaxFF~\cite{keith2010reactive}) account for bond directionality via an angular dependence. The existing set of parameters for these pre-defined models perform well for bulk structures, and close-packed (bulk-like) Au cluster configurations (at large sizes) but fail to describe non-compact Au cluster geometries (e.g., planar, hollow cages etc.). For instance, ReaxFF predicts the most stable isomer of Au$_8$ to be globular~\cite{keith2010reactive} in contrast to previous DFT calculations that show Au$_8$ to be planar.~\cite{xiao2006structural,serapian2013shape, goldsmith2019two} Given these challenges, gold catalytic clusters represent an excellent system for testing the efficacy of our AL scheme. We show that our AL-NN is able to adequately represent the energy landscape for diverse sizes and geometries as well as the dynamical properties of both clusters and bulk by sampling minimal amount of reference data ($\sim$ 500 total reference data).

\section{\label{sec:two}Methods}
\subsection{\label{sec:al} Active learning Scheme:}
Our AL strategy is shown schematically in Fig. \ref{fig:schematics} and involves the following major steps: (1) Training of the NN using the current structure pool (of Au nanoclusters configurations). (2) Running a series of stochastic algorithms to test the trained network’s current predictions. (3) An identification of configurational space where the NN is currently struggling. (4) An update of the structure pool with failed configurations. (5) Retraining of the NN with the updated pool and back to step 2. To test our AL scheme, we train a neural network to a reference DFT-PBE energetics for several gold configurations. The neural networks used in this study were constructed and trained using the Atomic Energy Network (AENet) software package,\cite{ARTRITH2016AENet} which was modified to implement the active learning scheme outlined above. Simulations using these networks were carried out using AENet interfaces with the Classy Monte Carlo simulation software\cite{ClassyMC} to perform the AL iterations. The main steps in our active learning iteration include:
\begin{figure}[ht]
\centering
\includegraphics[width =0.6\columnwidth]{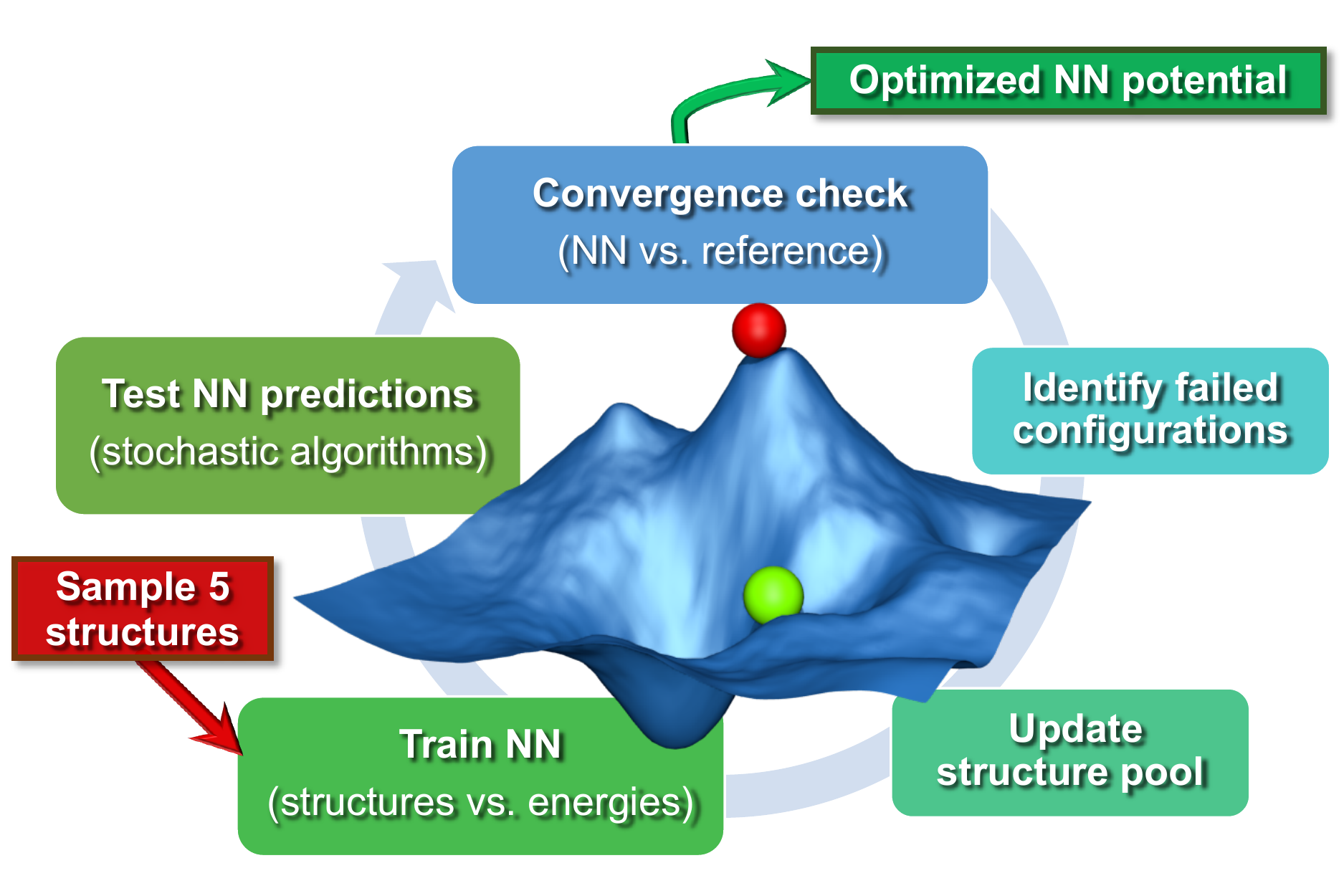}
\caption{Schematic showing the active learning workflow employed for generation of the NN potential model for Au nanoscale catalysts.}
\label{fig:schematics}
\end{figure}

\subsection{\label{sec:dft}DFT Calculations:} 
The Vienna Ab-initio Software Package (VASP)~\cite{kresse1996efficiency} with the Perdew-Burke-Eznerhof (PBE)~\cite{PhysRevB.54.11169} exchange-correlation functional was used to perform all the density functional theory calculations. The spin polarization was included in this DFT calculations. For element gold projector-augmented wave (PAW) potentials (PAW PBE Au 04Oct2007) provided with VASP were used. A single $k$-point at the center of the Brillouin zone was used for each calculation. Gaussian smearing with a width of 0.001 eV was used to set partial occupancies. The convergence criteria for the electronic self-consistent iteration and the ionic relaxation loop were set to be 0.1 meV and 1 meV per cluster, respectively. 

To evaluate the equation of state (EOS) plot for gold fcc gold lattice, $\pm$ 5\% strain was applied in all three directions. The initial bulk structure of gold fcc system has been collected from materials project database.~\cite{jain2013commentary} A dense $k$-point grid, defined by  $n_{atoms} \times n_{kpoints} \approx 1000$, where $n_{Atoms}$ is the number of atoms in the primitive cell and $n_{atoms} \times n_{kpoints} $ is the number of $k$-points were used in the DFT calculations for EOS plot.  A relatively high tolerance of $10^{-6}$ eV for energy convergence was employed in these calculations. Three independent elastic constants,  $n_{atoms} \times n_{Kpoints} \approx 1000$, was determined for cubic system by employing suitable lattice distortions~\cite{patil2006mechanical,manna2017tuning,wu2018characterization,manna2018large,manna2018enhanced, mckinney2018ionic,manna2019tuning2} represented by a strain tensor, $\varepsilon$, (defined by equation \ref{strain}) in such a way that the new lattice vectors $\bm{r}^{}$ in the distorted lattice is given by $\bm{r}^{'} = (\bm{I} + \varepsilon)\bm{r}$  where $\bm{I}$ is the unit matrix.
\begin{equation}\label{strain}
\varepsilon = \begin{pmatrix}
e_1 & \frac{e_6}{2} & \frac{e_5}{2} \\
\frac{e_6}{2} & e_2 & \frac{e_4}{2} \\
\frac{e_5}{2} & \frac{e_4}{2} & e_3
\end{pmatrix}
\end{equation} 
\begin{table}[htbp]
	\centering
	\caption{Three strain combinations in the strain tensor for calculating the three elastic constants ($C_{11}$, $C_{12}$, and $C_{44}$) of cubic structure of fcc gold. The magnitude of applied strain is varied in steps of 0.005 from $\delta$=-0.02 to 0.02.  $\Delta E$ is the difference in energy between that of the strained lattice and the unstrained lattice. The unstrained lattice volume is $V_0$.}
 	\label{table:elastic}
	\begin{tabular}{ l   c   c }
		\hline \hline
		Strain  &  Parameters (unlisted $e_i=0$) & $ \Delta E/V_{0}$ \\  
		\hline 
		1      &   $e_1=e_2=\delta, e_3=\frac{1}{(1+ \delta)^2}-1 $ & $3(C_{11}-C_{12})\delta^2 $ \\
		2      &   $e_1=e_2=e_3=\delta $ & $\frac{3}{2}(C_{11 + C_{12}})\delta^2$ \\  
        3      &   $e_6 =\delta, e_3 = \delta^2 (4-\delta^2)^{-1}$ & $\frac{1}{2}C_{44}\delta^2$ \\
	
		\hline \hline
	\end{tabular}
\end{table}
\subsection{\label{sec:NN}NN Architecture:}
Our NN consists of four layers of neurons; all the neurons/nodes of a layer are connected to every node in the next layer by weights in the manner of an acyclic graph. The two intermediate layers (hidden layers) consist of 10 nodes each. The input layer has 26 nodes which hold 26 symmetry functions that represent co-ordinates of the gold's potential energy surface (PES). The output layer consists of one node that represents the potential energy of a gold atom in a given configuration. Besides, the input layer and the hidden layers contain a bias node that provides a constant signal to all the nodes of its next layer. The choice of this network topology is based on a large number of trials for capturing various thermophysical properties of gold clusters. The Cartesian coordinates of a given gold atom are mapped into rotational and translational invariant co-ordinates as 
\begin{eqnarray}
G_{i}^{1} = \sum_{j} e^{-\eta(r_{ij}-R_{s})}\cdot f_{c}(r_{ij}) 
\end{eqnarray}
\begin{eqnarray}
\begin{aligned}
%\begin{align}
G_{i}^{2} = 2^{1-\zeta}\sum_{j, k\neq i}^{N} (1 + \lambda \cos \theta_{ijk})^{\zeta} \cdot e^{-\eta(r_{ij}^2+r_{ik}^{2})} \cdot f_{c} (r_{ij}) \cdot f_{c}(r_{ik}) 
%\end{align}
\end{aligned}
\end{eqnarray}
Here, $f_{c}(r_{ij}) = 0.5 [\cos (\frac{\pi r_{ij}} {R_{c}}) + 1]$ for $r_{ij} < R_{c}$ and $f_{c}(r_{ij}) = 0.0$ otherwise. In equations 2 and 3, the $r_{ij}$ corresponds to the distance between $i^{th}$ and $j^{th}$ particles of a gold cluster and $\theta_{ijk}$ is the angle formed by $r_{ij}$ and $r_{ik}$. The indices $i$, $j$ and $k$ run over all the particles in a cluster, which are with within  a cut-off distance $R_c= 6.0 \AA $. We have used 8 radial symmetry functions $G^1$ each with a distinct value of  $\eta$, which are tabulated in Table \ref{G2G4}. Similarly, 18 angular symmetry function  are used, each with a distinct set of  values. The parameters of these 18 angular symmetry functions are reported in Table \ref{G2G4}. The functional forms of these symmetry functions (Behler-Parrinello type symmetry functions \cite{behler2011atom}) have been used successfully to construct PES of different molecular systems, and thus adopted for this work. 

\begin{table*}[ht]
\caption{\label{G2G4} Parameters of the 8 radial symmetry functions $G^{1}$ and 18 angular symmetry functions $G^{2}$ with a cut-off distance of 6.0 \AA}
\includegraphics[width =0.9\columnwidth]{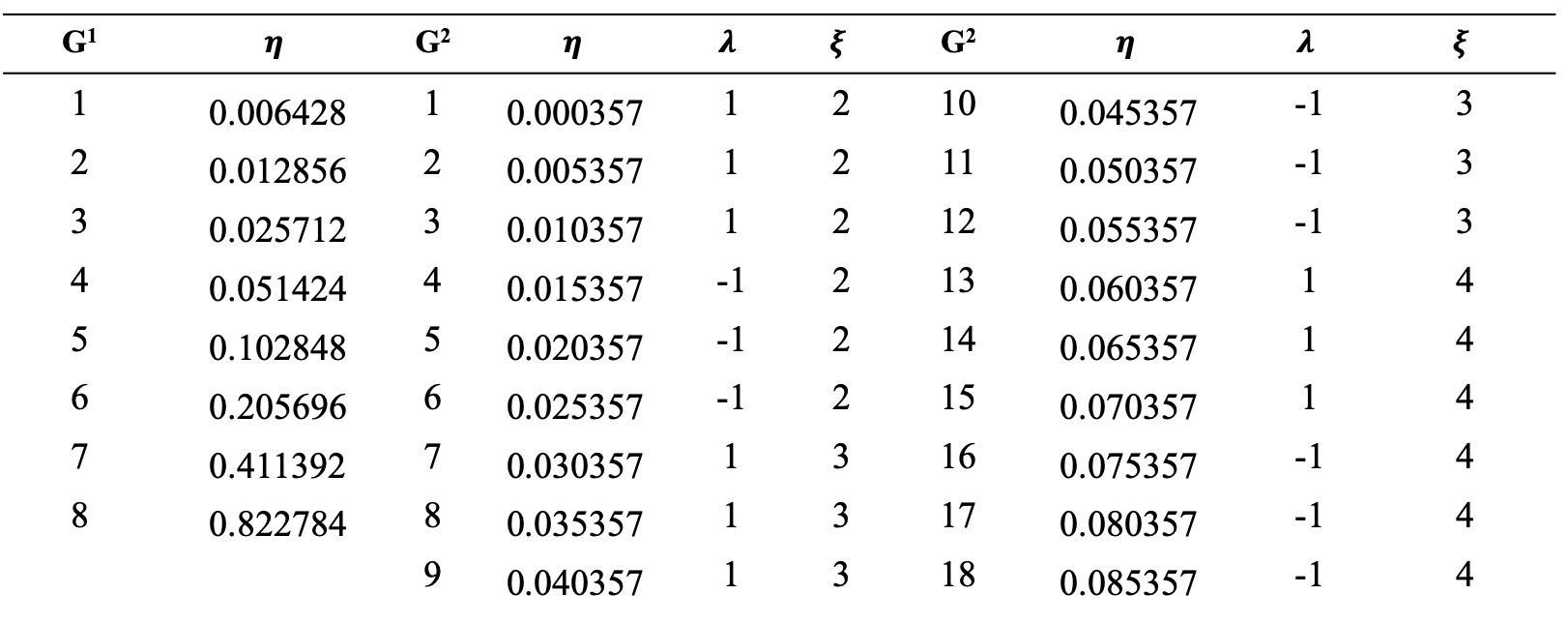}
\end{table*}

In this work, each and every atoms of a gold  cluster is represented by a NN and the total energy of the cluster is defined as   $E=\sum_{i}^{N_A} E_i$, where $E_i$ is the output of the $i^{th}$ NN, and NA is the total number of gold atoms in a given cluster which is same as the number of NNs. We note that the architecture and weight parameters of all these atomic NNs are identical. During the training, the symmetry functions of each atom of a configuration are fed to the corresponding NN via its input layer. In every NN, all the compute nodes in the hidden layers receive the weighted signals from all the nodes of its previous layer and feeds them forward to all the nodes of the next layer via an activation function as $x_{ij} = f (\sum_{k} W^{i}_{k, j} x_{(i-1), k})$. Here, $f(x) = \tanh(x)$  is used as the activation function of all the compute nodes. As mentioned earlier, the sum of all the outputs from all the NNs serves as the predicted energy of the system. The error in the NNs, which is the difference between the predicted and reference energies of a given configuration, is propagated backward via the standard back-propagation algorithm. All the weights that connect any two nodes are optimized using the Levenberg-Marquardt method\cite{pujol2007solution} in order to minimize the error, as implemented within the framework of AEnet\cite{ARTRITH2016AENet} open-source code.
\begin{figure*}[ht]
\centering
\includegraphics[width =\columnwidth]{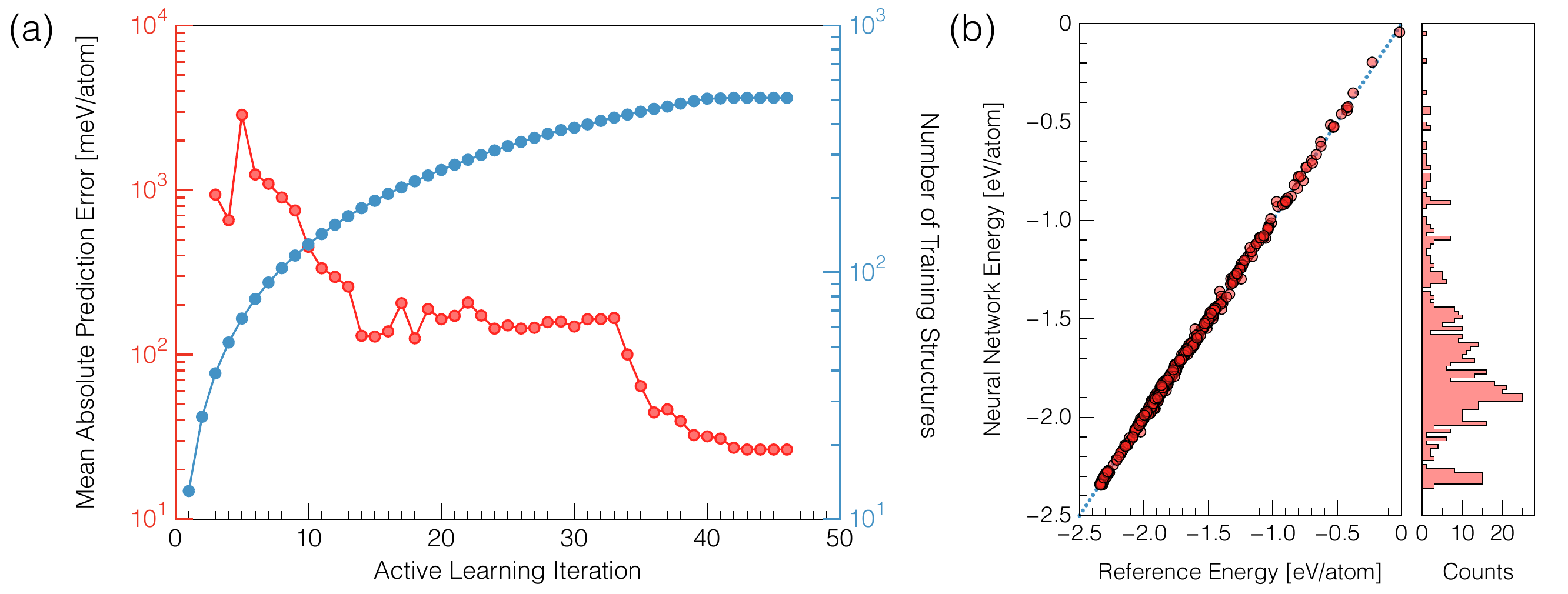}
\caption{Active learning of a NN potential for gold nanoclusters from sparse first-principles data. (a) The mean absolute error of the AL-NN tested on the DFT test set is plotted as a function of active learning iteration or generation (solid red dots). The scale on the RHS of the plot shows the size of the training data (solid blue dots) for the same training generation. (b) A correlation plot showing the performance of the final optimized network on the 579 structure training set.} 
\label{fig:al}
\end{figure*}
\subsection{NN training and optimization:}
A Levenberg-Marquardt approach\cite{singh2007ann} was used to optimize the neural network weights for each AL generation. This was done with a batch size of 32 structures and a learn rate of 0.1 once the structure pool was large enough to accommodate these settings. Initially, the batch size was set to 1, given the small initial training data set. For each network generation, the neural network is trained for a total of 2,000 epochs, where each epoch represents one complete training cycle. The AENet makes use of a $k$-fold cross validation scheme, where a given fraction ($k$) of the training set is not used for the objective minimization. Instead this fraction is used to cross validate the training process to minimize over-fitting. For each AL iteration, the network which had the best error from the cross validation was chosen as the best network for this AL iteration and is carried forward.

\subsection{Configuration Sampling:}
Once the best network has been chosen, a series of simulations are run to actively sample the configurational space predicted by the current NN. It was found that MD is not suitable for sampling within this scheme due to the fact that when the network is still in its infancy, large spikes in the forces can lead to unphysical acceleration of particles within the simulation box. In addition, even in a reasonably well-trained network, MD can be trapped in a local energy well that prevents it from searching the phase space outside of this well. This can often create models that work well within the trained local minima, but can have catastrophically bad predictions when the model is applied to environments found outside of the training set. Monte Carlo and other similar sampling methods in contrast are much less sensitive to spikes in the energy surface which make them more suitable methods for sampling poorly trained energy landscapes.

In addition, a wide collection of non-physical moves or non-thermal sampling approaches can be used. For the purposes of this work, Boltzmann based Metropolis sampling and a nested ensemble based approach~\cite{nielsen2013nested} were used to generate the structures for each AL iteration. This was done to gather information on both thermally relevant structures predicted by the neural network as well as higher energy structures which may still be important for creating an accurate model. The Metropolis simulation was run for 5,000 MC cycles at 300K with the initial structure being randomly picked from the current neural network training pool. The Nested Ensemble simulations were run for another 5,000 cycles.

\begin{figure*}[ht]
\centering
\includegraphics[width =0.9\columnwidth]{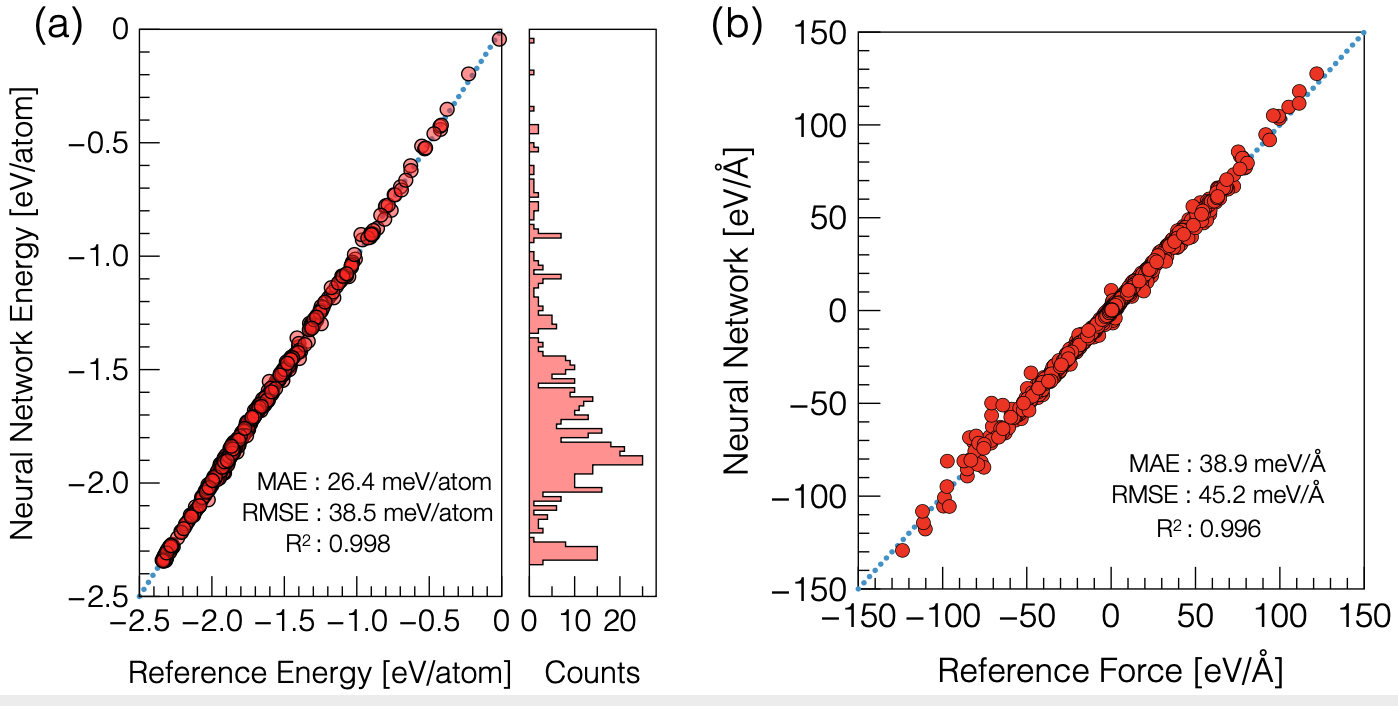}
\caption{Performance of the actively learned NN model on an extensively sampled test data set. Energy correlations comparing actively learnt NN-prediction with the reference DFT energies for a test set that comprises of $\sim{~}$ 1100 Au cluster configurations. The dotted blue line represents the zero MAE. (b) Force correlations comparing actual force and that analytically derived from the actively learnt NN over the same test set. The calculated MAE, RMSE and R$^2$ are provided in the inset of the plot.}
\label{fig:mae}
\end{figure*}

\subsection{Testing of the NN}
After the stochastic sampling step is completed, a set of 10 structures are gathered from the trajectory files of the Metropolis and Nested Sampling files. These are sampled by outputting a structure every 1000 number of cycles for both the nested ensemble run and metropolis. For the nested sampling run this is set up such that we pull one structure from each energy ``strata" as the nested sampling gradually constricts the energy space. This ensures we are always testing structures from both high energy and low energy regions of the phase space. The real energy of these structures are computed using DFT-PBE and compared with the predictions of the NN model. For each structure, if the neural network and the DFT prediction do not agree within a given tolerance, the structure is then added to the training pool to be used for the next AL iteration. This entire process is continued until the exit criteria is hit. For this work, we specified that if no new structures were added in 5 consecutive AL iterations, that the potential has converged. For the addition tolerance, we specified that any structure with a greater than difference of 20 meV between the real and predicted energy should be added to the training pool. The acceptable tolerance is based on typical prediction errors of DFT (the reference model) which is around 20 meV.~\cite{jain2011high} Also, the kT value for room temperature is $\sim$ 25 meV - so the errors are within typical thermal fluctuations at room temperature.
\begin{figure*}[ht]
\centering
\includegraphics[width =0.7\columnwidth]{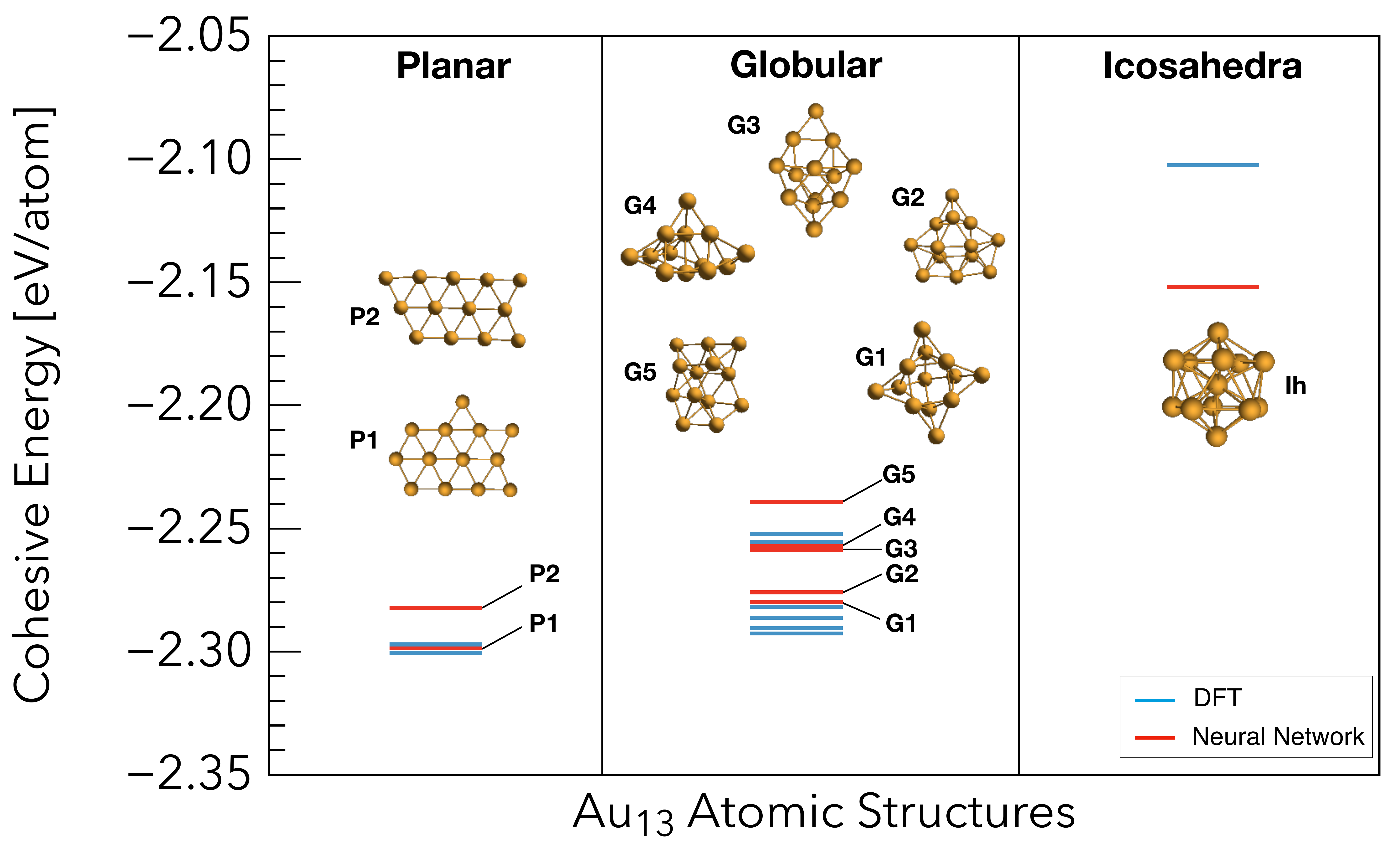}
\caption{Predictive power of the AL-NN for the various 2D and 3D Au$_{13}$ configurations with respect to the DFT-predicted global energy minimum structure. The cohesive energies of planar structures, intermediate configurations and 3D icosahedron (Ih) computed with AL-NN are compared with those obtained by DFT. The blue and red solid lines correspond to the DFT predicted and Neural Network predicted cohesive energies respectively.} Most of the available EFFs predict the globular Ih to be the most stable structure for Au$_{13}$ in contrast to DFT (which predicts planar to be the global energy minimum). AL-NN describes the energetics of Au$_{13}$ clusters in excellent agreement with DFT calculations. 
\label{fig:predict13}
\end{figure*}
\subsection{Initialization of the AL-NN}
The initial neural network cannot be trained on zero data, a single structure is used to seed the initial neural network in order to kick off the training process. This was chosen to be a reasonably minimized structure in order to ensure at least one low energy configuration was contained in the training set. Theoretically one could begin with any number of seed structures, but for the purposes of evaluating the efficiency of this approach, the absolute minimal seed data was used. In order to rigorously validate the neural network models, we created a test set that consists of roughly $\sim$ 500 configurations of Au clusters. 

\begin{figure*}[ht]
\centering
\includegraphics[width =0.7\columnwidth]{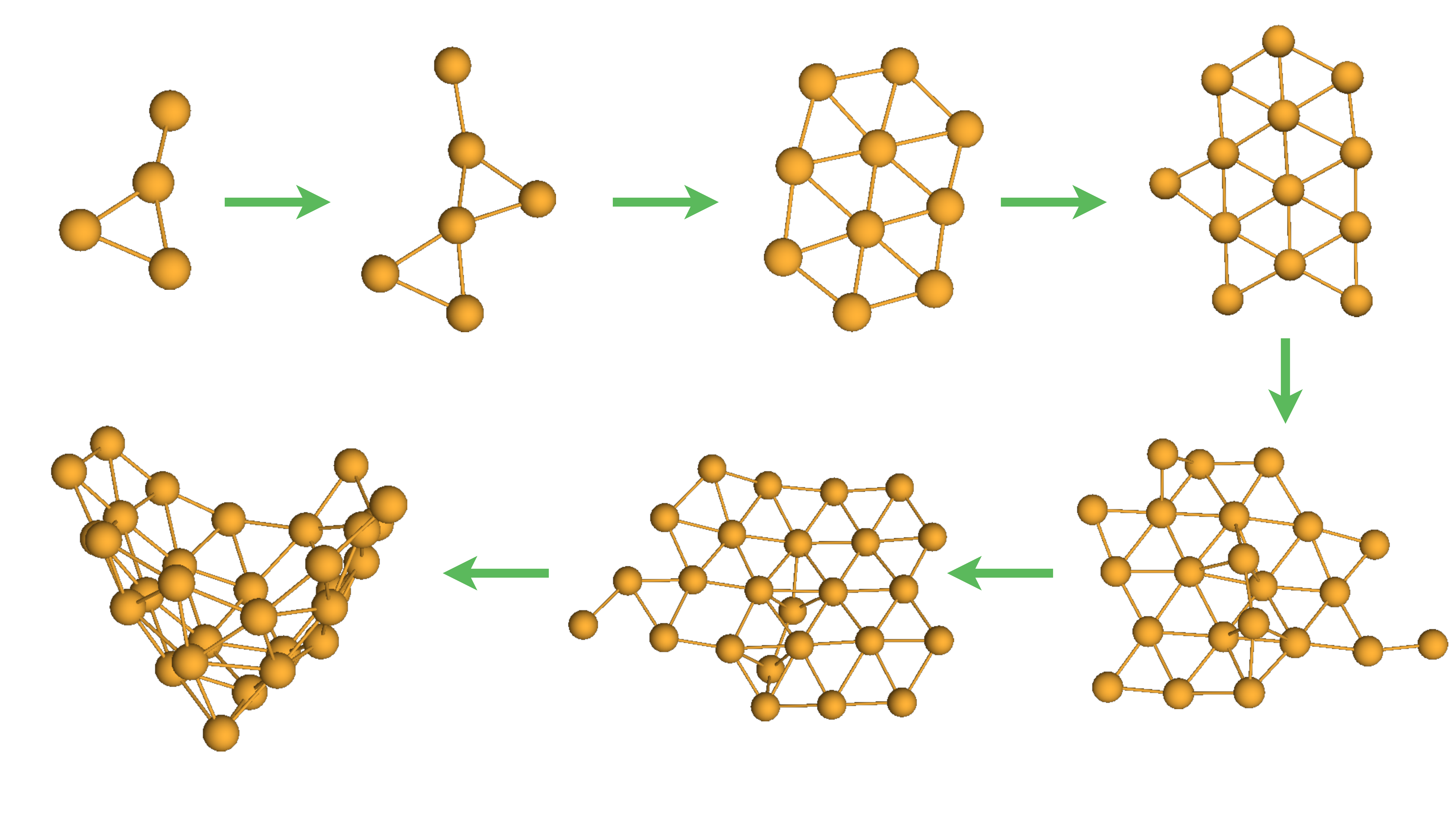}
\caption{The size-dependent transition of a gold cluster at 300 K as predicted by the NN developed in this study. The atoms are introduced one-by-one into the system via grand canonical swap moves.  The Monte Carlo simulation was run for a total of 100,000 cycles at 300K and a constant gas phase reservoir density.}
\label{fig:dynamics}
\end{figure*}

\begin{table*}[htbp]
	\centering
	\caption{Structural, energetic, and elastic properties of bulk polymorphs of gold as predicted by the AL-NN model developed in this study. These predictions are compared with values obtained from our DFT calculations, and previous experiments (if available). $E_C^{fcc}$ refers to cohesive energy of FCC, $a_j$ to lattice parameter of cubic polymorph $j$, and  $\Delta E^{j-fcc}$ is the difference of cohesive energy between polymorph $j$ and FCC. The quantities $C_{ij}$ are the values of elastic stiffness constants.}
\label{table:bulkAu}
\begin{adjustbox}{width=\columnwidth,center}
\begin{tabular}{lccccccc}
\hline\hline
\\
               & Neural Network & EAM~\cite{foiles1986embedded} & SC\cite{sutton1990long}  & ReaxFF\cite{keith2010reactive} & HyBOP\cite{narayanan2016describing} & DFT & Experiment \\
               & (This Study) &  &   &  &  & (This study) & \\
\hline               
$E_C^{fcc}$ (eV/atom) & -3.22                     & -3.93        & -3.78       & -3.77           & -3.82         & -3.22          & -3.81\cite{foiles1986embedded}      \\
$a^{fcc}$ (\AA)        & 4.15                       & 4.08         & 4.08        & 4.18            & 4.19          & 4.17            & 4.07\cite{foiles1986embedded}       \\
$\Delta E^{bcc-fcc}$ (eV/atom)     &       0.02                     & 0.02         & 0.03        & 0.14            & 0.08          & 0.02            & -          \\
$a^{bcc} $ (\AA)        &       3.306                     & 3.24         & 3.25        & 3.31            & 3.32          & 3.31            & -          \\
$\Delta E ^{sc-fcc}$ (eV/atom)     &       0.32                     & 0.39         & 0.28        & 0.69            & 0.5           & 0.20            & -          \\
$a^{sc}$ (\AA)         &   3.08                         & 2.65         & 2.72        & 2.95            & 2.82          & 2.76            & -          \\
$\Delta E^{dia-fcc}$ (eV/atom)   &   1.22   & 0.94         & 0.60        & 1.0             & 1.37          & 0.71            & -          \\
$a^{dia}$ (\AA)        &       6.83                     & 5.75         & 6.07        & 6.72            & 6.45          & 6.18            & -          \\
$C_{11}$ (GPa)            & 171                      & 183          & 180         & 168             & 231           & 150             & 192\cite{cagin1999calculation}        \\
$C_{12}$ (GPa)            & 157                      & 159          & 148         & 130             & 170           & 129             & 163\cite{cagin1999calculation}        \\
$C_{44}$ (GPa)       & 42                       & 45           & 42          & 55              & 75            & 31              & 42\cite{cagin1999calculation}        \\
\hline\hline
\end{tabular}
\end{adjustbox}
\end{table*}

\begin{figure}[ht]
\centering
\includegraphics[width =0.5\columnwidth]{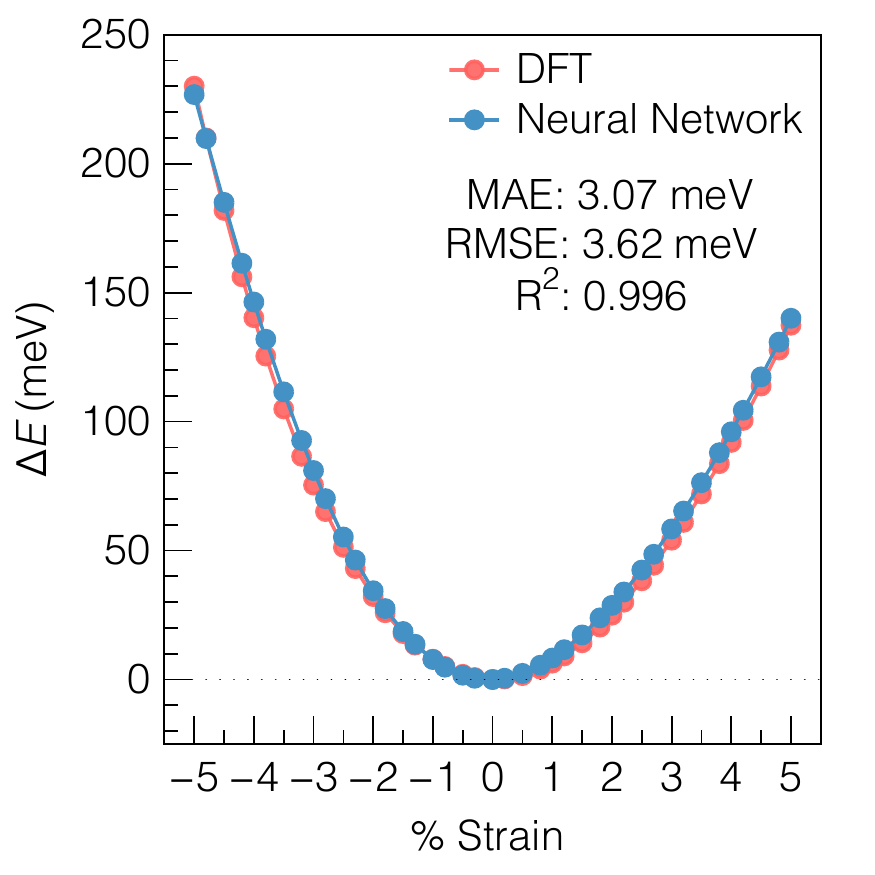}
\caption{Comparison of the AL-NN equation of state with that obtained from DFT. The calculated MAE, RMSE and R$^2$ are provided in the inset of the plot.} 
\label{fig:eos}
\end{figure}

\section{Results}
First, we evaluate the performance of our active learning (AL) scheme depicted in Fig. \ref{fig:schematics}. Fig. \ref{fig:mae} shows the mean absolute error (MAE) in meV/atom as a function of epochs – each epoch is an AL iteration or a complete training cycle. Fig. \ref{fig:mae}(a) also shows the number of structures added during each of the AL iteration. Our training of the gold NN is initiated with minimalist number of training configurations. Therefore, the initial NN have very high errors $\sim$ 1000-3000 meV/atom. With each AL epoch or iteration, the NN learns and begins to sample configurations in regions of failure. We note that the errors (NN vs. DFT predictions) drop progressively from $\sim$ 1000 meV/atom at iteration 1 to $\sim$ 20 meV/atom at $\sim$ 50 epoch. Initial training errors are nearly on the same magnitude as the total system energy. The NN learns rapidly in the beginning as more distinct (failed) cluster configurations are added to the pool. The MAE drops sharply and plateaus at AL iteration $\sim$ 10 suggestive that the NN search is stuck at a local minimum. After about a total of 50 AL epochs or iterations, we see that the MAE drops to  $\sim$ 20 meV/atom. At this point, the gold NN reaches our prescribed stopping criteria i.e. no new structures are added during 5 consecutive test cycles. It is worth noting that final structure count at this point has reached a total of  $\sim$ 500 unique training structures. Fig. \ref{fig:al} plots the correlation between the performance of the final optimized NN trained on the  $\sim$ 579 configuration training set. The AL-NN predictions of energetics for the clusters in the training pool are compared with that the reference DFT model. MAE for the training set was found to be less than 20 meV/atom, which is of the same order of magnitude as DFT error. 

We next evaluate the network performance as a function of the number of AL epochs. We choose the best network from each AL iteration and test its performance on a test set that comprises of 1101 configurations and their energies. Fig. \ref{fig:al} (a) shows the correlation between AL-NN predicted energies vs. reference DFT-PBE energies.  As expected, we find that the final optimized NN is able to reliably predict the Au cluster energies for an elaborate test data set generated not only near equilibrium, but also in the highly non-equilibrium region that extends far beyond. As a more rigorous test of the performance of AL-NN, we compute the forces on the atoms for the various clusters and compare those with that obtained from DFT-PBE. It should be noted that the forces were not included as part of the training during the AL iterations. Fig. \ref{fig:al} (b) shows the correlation between the AL-NN predicted vs. DFT forces. Each point in this correlation plot represents one of the force components - $F_x$, $F_y$ and $F_z$ acting on a particle. We find the overall MAE between AL-NL vs. DFT predicted forces is $\sim$ 20 meV/\AA. Given that the NN had not been trained on the forces, this agreement with the DFT is of excellent quality.  Overall, our Au AL-NN optimized network performs very well over an extensively sampled test data set.

We also test the performance of AL-NN in reproducing the DFT predicted energetic ordering of structural isomers at a given cluster size. Fig. \ref{fig:predict13} compares the energetic ordering of representative planar, intermediate and globular isomers of Au$_{13}$ clusters predicted by AL-NN with those from DFT calculations (atomic coordinates were relaxed at the corresponding level of theory). The Au$_{13}$ isomers depicted are configurations that lie near the global energy-minimum as predicted by DFT. The AL-NN predict the correct energetic ordering consistent with the DFT predictions although it should be noted that the energetic difference between various isomers are slightly underpredicted. This performance is still much better than those of most popular existing empirical force-fields such as EAM and its variants which incorrectly predict the icosahedral structure to be more stable than planar for $Au_{13}$.

A crucial yet challenging test of any FF is its ability to accurately capture the size dependent global minimum energy (GM) configuration especially for cluster sizes that are not part of the training set. Such as test can be regarded as a true test of its transferability. The AL-NN is successful in predicting GM configurations for nanoclusters at several sizes as shown in Fig. \ref{fig:dynamics}. In accordance with the DFT predictions, our AL-NN GM structures are planar for clusters with size $n < 14$ atoms and globular at higher sizes. Our predicted critical size for planar to globular transition (13 atoms) is identical to previous DFT calculations,~\cite{kinaci2016unraveling, narayanan2016describing} and previously reported ion mobility and spectroscopy measurements.~\cite{collings1994optical} AL-NN predicted planar GM structures for sizes up to 13 atoms match the ones reported in Ref.~\cite{gruber2008first,wang2002density} using DFT calculations. AL-NN also successfully reproduce the evolution of various structural motifs with cluster size in accordance with DFT predictions and spectroscopic experimental observations.~\cite{gruber2008first,wang2002density,collings1994optical}

While the snapshot energies show that the examined planar structures are lower in energy than the globular counter parts, an actual molecular simulation is required to check for other unknown states that might potentially be lower. To examine this a Grand Canonical Ensemble Monte Carlo simulation was performed using the Aggregation Volume Bias Monte Carlo approach. In these simulations the cluster is slowly grown atom by atom along with standard Monte Carlo moves in order to observe how the cluster configures itself under thermal motion. The snapshot results of these simulations can be found in Fig. \ref{fig:dynamics}. In good agreement with the snapshot data, the cluster stably grows in a planar configuration up to 13 atoms in size. As the cluster growth further the addition of atoms onto the top of the 2D structure can be observed.  Overtime more and more atoms are added in out of plane positions and the cluster begins to fold in on itself and begins to form a cage structure. Our simulations predicts the correct trend in the geometry as a function of cluster size.  Because of the statistical improbability of forming 2D clusters using 3D Monte Carlo moves and given that 3D is more entropically favored than 2D, we can safely conclude that planar structures are structurally stable under thermal conditions.

Apart from describing the clusters accurately, it is also essential for the AL-NN to appropriately describe the bulk as many fundamental research problems including adsorption on Au surfaces, diffusion of clusters on surfaces of bulk Au, and breakdown of large Au clusters into small ones upon high energy impact. Fig. \ref{fig:eos} shows the comparisons of the equation for state for bulk Au compared to that obtained from DFT. We note that the agreement between the AL-NN predictions and DFT is excellent. The structural, elastic, and cohesive energies of various bulk Au polymorphs predicted AL-NN with those from DFT/experiments are summarized in Table \ref{table:bulkAu}. AL-NN preserves the DFT evaluated energetic ordering of the various bulk polymorphs and predicts the FCC to be the most stable bulk polymorph in agreement with previous DFT and experiments. AL-NN also predicts lattice parameter of FCC Au to be 4.15 \AA, in good agreement with our DFT calculations (4.17 \AA), and previous experiments (4.07 \AA).\cite{foiles1986embedded} The cohesive energy for FCC Au predicted by AL-NN (-3.22 eV/atom) agrees very well with previous experiments (-3.81 eV/atom).\cite{foiles1986embedded} Note that DFT-PBE significantly underestimates this value (-2.97 eV) due to its inadequate treatment of the dispersion effects in Au.\cite{serapian2013shape, kinaci2016unraveling} AL-NN predictions for the elastic constants are in excellent agreement with experiments as well as spherically symmetric potentials (i.e., EAM, SC). In particular, one of the challenging elastic properties to describe is the ratio $C_{12}/C_{44}$. Our AL-NN predicts this ratio to be 3.78 in excellent agreement with experimental value of 3.9 and DFT value of 4.1. EAM and SC perform well for this ratio giving value $\sim$ 3.5. Considering that the AL-NN potential was not fitted explicitly to bulk properties, the agreement with DFT values and experiments is quite excellent.

\section{Discussion}
The AL-NN training approach has shown it is capable of creating a NN that does an excellent job of replicating DFT energies using relatively small training data sets. Only performing on the order of a few hundred DFT calculations, we build an NN model that was not only able to capture the cluster properties but also perform well on bulk test set (despite the bulk properties not being included in the training data).

\begin{figure}[ht]
\centering
\includegraphics[width =0.55\columnwidth]{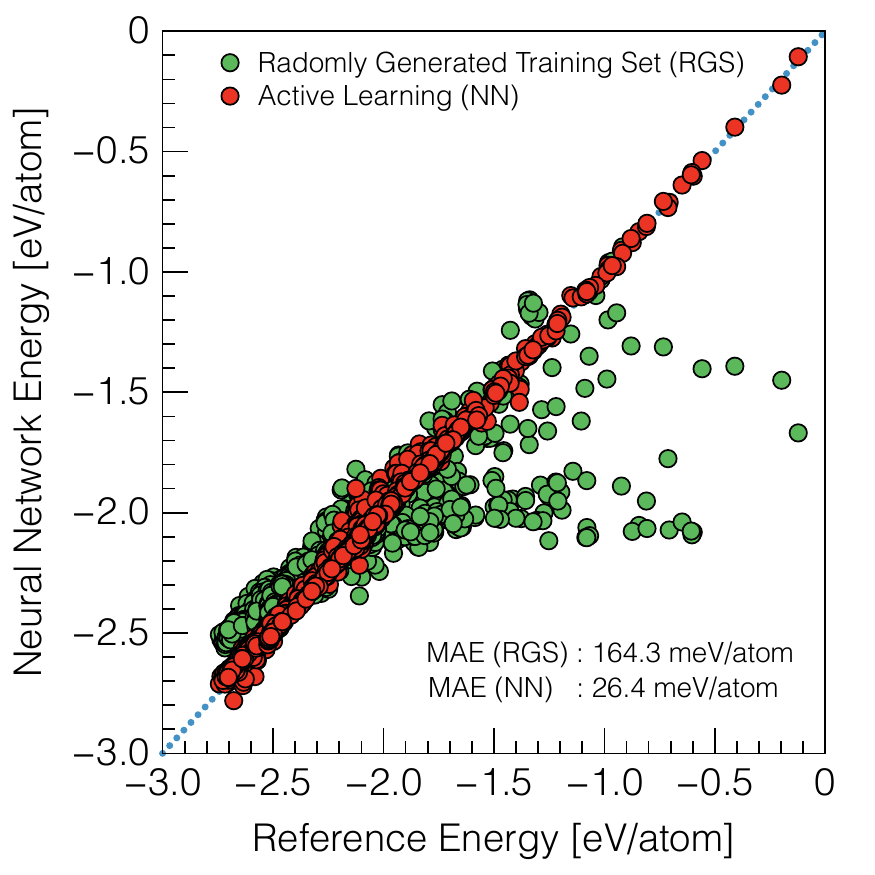}
\caption{Energy correlation plot for randomly generated training set vs. AL generated training set. The mean absolute error (MAE) are provided in the inset.} 
\label{fig:AlvsRandom}
\end{figure}

\begin{figure}[ht]
\centering
\includegraphics[width =0.75\columnwidth]{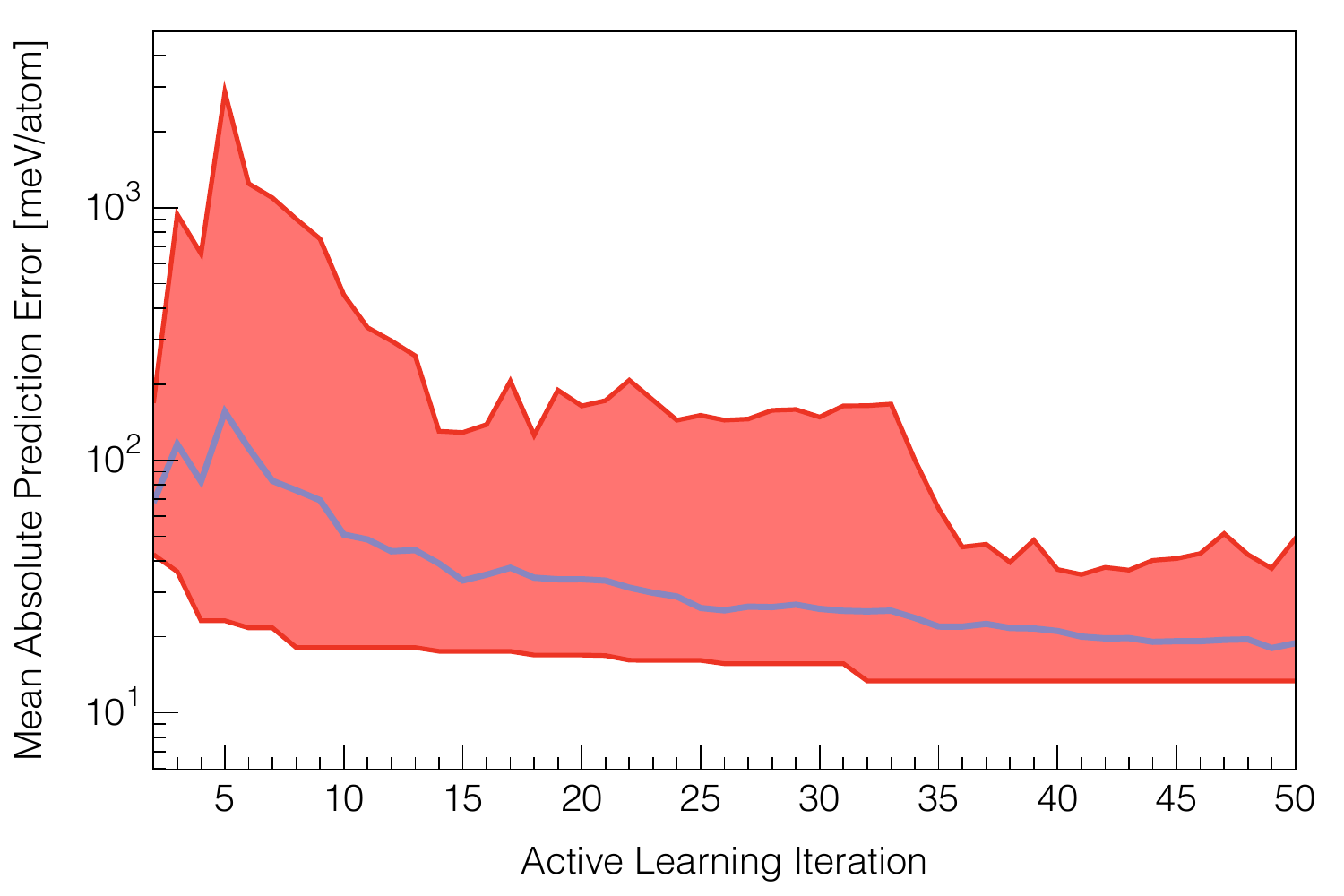}
\caption{Average MAE error against the validation set as a function of AL iteration (blue curve) and the error bounds (red region).} 
\label{fig:repro}
\end{figure}

We next test the effectiveness of the nested ensemble sampling scheme vs. a random sampling (see Fig. \ref{fig:AlvsRandom}). For comparison, we generated a neural network whose training set was created by a mixture of random structure generation and thermal sampling using a MD force-field. An identical number of structures were randomly sampled similar to that during iterative AL runs and used to train a neural network. The Randomly Generated Structure (RGS) approach created a network that had a MAE of 164.3 meV/atom error with many of the energies grossly over predicted compared to the reference DFT model.  On the other hand, AL approach has an MAE of 26.4 meV/atom for the same number of training structures. This clearly shows that the nested ensemble search performs much better compared to a random search.

We next assess the reproducibility as well as the influence of the starting configurations on the NN evolution during the AL cycle. To show the reproducibility of the AL algorithm, we perform an ensemble of 30 AL runs.  The results are plotted in Fig. \ref{fig:repro}  which shows the average MAE error against the validation set as a function of AL iteration (blue curve). The error bounds are given as the worst case (lower) and the best case (upper) scenario of the 30 independent AL runs.  All 30 runs converge after $\sim{~}$ 35 AL iterations - the worst network has $\sim{~}$ 35 meV/atom error and the best network $\sim{~}$ 20 meV/atom. This clearly shows that regardless of the starting configuration, the AL scheme converges to an optimal network after sampling a few hundred Au cluster configurations.

Finally, there are at least two main computational cost advantages to our AL procedure. First, the total cost for training the AL-NN with a training data size increasing up to 500 Au cluster configuration is approx. 30-40 core-hrs. The same for training a conventional NN with $\sim{~}$100,000 cluster configuration is $\sim{~}$90-120 core-hrs for convergence to solution. Secondly and more importantly, the major gain is in the generation of the training data. For the reference DFT calculation of energies for a 30 atom Au cluster, each single shot energy calculation requires $\sim{~}$240 core-hrs. Given that the training data size for a conventional NN is $\sim{~}$200X more compared to that in AL, this translates into a cost saving of $\sim{~}$48,000 core-hrs for the AL-NN compared to traditional NN. The extent of cost savings for the AL-NN depends on the fidelity of the quantum calculation as well as the cluster sizes being sampled. The computational cost savings are clearly much more for higher fidelity calculations such as CCSD and QMC and for larger cluster sizes in the training data.

\section{Conclusions}
In summary, we introduce an automated active learning workflow for building NN models against a sparse first-principles training dataset constructed on-the-fly. Our AL scheme allows for on-the-fly sampling of both the configurational and potential energy surface of gold nanoclusters of different sizes and builds a high-quality neural network with a sparse dataset that comprises of only $\sim$ 500 reference DFT evaluations. Using an extensive DFT test set of 1101 configurations, we show that our NN provides excellent predictions of both the energies and forces over a wide variety of cluster sizes (that were not originally part of the training set). Our AL trained NN captures the global minimum energy configurations from several different samples of cluster sizes and the energetic ordering i.e. stability of various cluster configurations at any size. It also captures the size dependent critical size of transition from planar to globular clusters consistent with DFT calculations. The NN also predicts the evolution of structural motifs with cluster size. Moreover, it also reasonably captures the thermodynamics, structure, elastic properties, and energetic ordering of bulk condensed phases, in excellent agreement with DFT calculations and previously reported spectroscopic experiments. Given that high-fidelity quantum calculations such as quantum Monte Carlo (QMC)\cite{vincent2007quantum} and coupled clusters (CCSD)\cite{scuseria1989coupled} are computationally expensive and hence, even with the significant improvements in computing resources, one can only generate sparse data sets. From this perspective, our AL scheme overcomes a major limitation of training NN against sparse datasets. Finally, our work lays the groundwork for future construction of NN to describe the complex potential energy landscape and dynamics of catalytic nanoclusters by training against sparse high-fidelity data obtained from minimal quantum calculations. 

\textit{Acknowledgement} We acknowledge funding from BES Award DE-SC0020201 by DOE to support this research. The use of the Center for Nanoscale Materials, an Office of Science user facility, was supported by the U.S. Department of Energy, Office of Science, Office of Basic Energy Sciences, under Contract No. DE-AC02- 06CH11357. This research used resources of the National Energy Research Scientific Computing Center, which was supported by the Office of Science of the U.S. Department of Energy under Contract No. DE-AC02-05CH11231. An award of computer time was provided by the Innovative and Novel Computational Impact on Theory and Experiment (INCITE) program of the Argonne Leadership Computing Facility at the Argonne National Laboratory, which was supported by the Office of Science of the U.S. Department of Energy under Contract No. DE-AC02-06CH11357. SKRS acknowledges UIC start-up funds for supporting this research. 

\bibliography{ref}
\end{document}